\documentclass[useAMS,usenatbib]{mn2e}
\usepackage{graphicx}
\usepackage{pslatex}
\usepackage{natbib}
\usepackage{txfonts}
\newcommand{\Eqref}[1]{(\ref{#1})}

\newcommand{\cH}{\mathcal{H}}
\newcommand{\cK}{\mathcal{K}}

\newcommand\Chi{{(\chi^2_\nu)^{1/2}}}

\def\idm#1{{\mbox{\scriptsize #1}}}
\renewcommand{\vec}[1]{\mathbf{#1}}

\begin{document}

\title{The long-term stability of extrasolar system HD~37124.
Numerical study of resonance effects.
}
\author[Go\'zdziewski, Breiter and Borczyk]
{Krzysztof Go\'zdziewski$^1$\thanks{Toru\'n Centre for Astronomy,
N. Copernicus University, Poland, k.gozdziewski@astri.uni.torun.pl},
S\l awomir Breiter$^2$\thanks{Astronomical Observatory of A. Mickiewicz University,
 S\l oneczna 36, PL 60-286 Pozna\'n, Poland, breiter@amu.edu.pl},
Wojciech Borczyk$^2$\thanks{Astronomical Observatory of A. Mickiewicz University,
 S\l oneczna 36, PL 60-286 Pozna\'n, Poland, bori@moon.astro.amu.edu.pl}
 }
\maketitle
   \begin{abstract}
 We describe numerical tools  for the  stability analysis of extrasolar
 planetary systems. In particular, we consider the relative Poincar\'e variables
 and symplectic integration of the equations of motion. We apply the tangent map
 to derive a numerically efficient algorithm of the fast indicator MEGNO (a
 measure of the maximal Lyapunov exponent) that helps to distinguish  chaotic
 and regular configurations. The results concerning the three-planet extrasolar
 system HD~37124 are presented and discussed. The best fit solutions found in
 earlier works are studied more closely. The system involves Jovian planets with
 similar masses. The orbits have moderate eccentricities, nevertheless the best
 fit solutions are found in dynamically active region of the phase space. The
 long term stability of the system is determined by a net of low-order two-body
 and three-body mean motion resonances. In particular, the three-body resonances
 may induce strong chaos that leads to self-destruction of the system after Myrs
 of apparently stable and bounded evolution. In such a case, numerically
 efficient dynamical maps are useful to resolve the fine structure of the phase
 space and to identify the sources of unstable behavior.
   \end{abstract}

\section{Introduction}

{Understanding
}
the extrasolar planetary systems has became a major challenge for contemporary
astronomy.  One of the most difficult problems in this field concerns the
orbital stability of such systems.  Usually, the investigations of long-term
evolution are the domain of direct, numerical integrations. The stability of
extrasolar systems is often understood in terms of the Lagrange definition
implying that orbits remain well bounded over an arbitrarily long time. Other
definitions may be formulated as well, like the astronomical stability
\citep{Lissauer1999} requiring that the system persists over a very long, Gyr
time-scale, or Hill stability \citep{Szebehely1984} that requires the constant
ordering of the planets. In our studies, we prefer a more formal and stringent
approach related to the fundamental Kolmogorov-Arnold-Theorem (KAM), see
\cite{Arnold1978}. Planetary systems, involving a dominant mass of the parent
star  and significantly smaller planetary masses,  are well modeled by
close-to-integrable, Hamiltonian dynamical systems. It is well known, that their
evolution may be quasi-periodic (with a discrete number of fundamental
frequencies, forever stable),  periodic (or resonant; stable or unstable) or
chaotic (with a continuous spectrum of frequencies, and unstable). In the last
case, initially close phase trajectories diverge exponentially, i.e., their
Maximum  Lyapunov Characteristic Exponent (MLCE, denoted also with $\sigma$) is
positive. In general, the distinction between regular and chaotic trajectories
is a very difficult task that may be resolved only with numerical methods
relying on efficient and accurate integrators of the equations of motion.

The main motivation of this paper is to describe numerical tools that are useful
for studies of the dynamical stability and to apply them to the HD~37124 system
\citep{Vogt2005}. We recall the fundamentals of relative canonical  Poincar\'e
variables as  -- in our opinion -- one of the best frameworks for symplectic
integrators. These {canonical} variables are well suited for the
construction of a \citet{LasRob} composition method that improves a classical
Wisdom-Holman (W-H) algorithm \citep{WH:91}. We supplement the integrator with a
propagator of the associated symplectic tangent map that approximates the
solution of variational equations \citep{MikIn:99}. Finally, we compare two fast
indicators that reveal the character of phase trajectories. The first one is a
relatively simple method for resolving fundamental frequencies and spectral
properties of a close-to-integrable Hamiltonian system -- a so called Spectral
Number (SN), invented by \citet{Michtchenko2001}. The second indicator belongs
to the realm of the Lyapunov exponent based algorithms; we chose the numerical
tool developed by \cite{Cincotta2000,Cincotta2003} under the name of MEGNO. In
this work, we refine the algorithm of MEGNO that makes explicit use of the
symplectic tangent map \citep{Gozdziewski2003b}.

As a non-trivial application of the presented numerical tools,  we consider the
3-planet system hosted by the HD~37124~star \citep{Vogt2005}. It has been
discovered by the radial velocity (RV) technique. The recent model of the RV
observations of HD~37124 predicts three equal Jovian type planets with masses
$\sim 0.6$~m$_{\idm{J}}$ in orbits with moderate eccentricities. In such a case,
the application of symplectic integrators without regularization is particularly
advantageous thanks to the numerical efficiency (long time-steps) and accuracy
(the total energy does not have a secular error and the angular momentum
integral is conserved). The number of multi-planet systems resembling the
architecture of the Solar system increases\footnote{For a recent statistics of
the discoveries, see Jean Schneider's Extrasolar Planets Encyclopedia,
http://exoplanets.eu.}. Hence, our approach may be useful in other cases.

\section{Numerical tools}

According to the classical results of celestial mechanics, the $N$-body problem
has only 10 integrals of motion for all $N > 2$; they consist of 6 integrals of
barycenter, 3 integrals of angular momentum and the energy integral. The
integrals of barycenter play a very particular role in the studies of an
$N$-body system dynamics. First, they define the origin of an inertial reference
frame in terms of the mutual distances and velocities of the bodies considered,
thus dismissing the need of some extrinsic absolute frame. But what is more
important, being linear forms of coordinates and momenta they allow a unique
reduction of the system, lowering the number of degrees of freedom by three,
with no loss of information. This is why we can solve the relative two-body
problem and then recover the motion of both masses with respect to their center
of mass. And this is why we can approximately solve the heliocentric motion of
planets, recovering the barycentric evolution \textit{a posteriori}.

Within the framework of Hamiltonian mechanics, the reduction is usually achieved
by means of a transformation to one of the two common variable types: relative
Jacobi variables, or ''heliocentric'' Poincar\'e variables
\citep[Ch.~XIII]{Whittaker}. We focus on the latter set, because it offers the
best choice in many aspects. We introduce the basic ideas related to the
Poincar\'e variables and we derive them as a Mathieu transformation; this way is
simpler and more intuitive than the procedure based upon a generating function
that was presented by \citet[p.~343]{Whittaker} or \citet{DLL:98}.
{
Then we discuss the setup of the Hamiltonian within the framework
of Wisdom-Holman type integrators.
}

\subsection{Poincar\'e variables basics}

Let us consider a system consisting of $N+1$ material points with masses $m_0,
\ldots ,m_N$. We define a barycentric position  vector $\vec{p} \in
\mathbb{R}^{3N}$ and its canonical conjugate momentum $\vec{P} \in
\mathbb{R}^{3N}$ as
\begin{equation}
\label{pP}
    \vec{p}=\left[ \begin{array}{c}
                     \vec{p}_0 \\
                     \vec{p}_1 \\
                     \vdots \\
                     \vec{p}_N \\
                   \end{array}
                   \right], \qquad
     \vec{P}=\left[ \begin{array}{c}
                     \vec{P}_0 \\
                     \vec{P}_1 \\
                     \vdots \\
                     \vec{P}_N \\
                   \end{array}
                   \right].
\end{equation}
Then, the barycentric equations
of motion can be derived from the Hamiltonian function
\citep{Laskar:90}
\begin{equation}\label{Hbar}
     \cH(\vec{p},\vec{P}) =
     \sum_{i=0}^N \frac{\vec{P}_i^2}{2\,m_i}
     -  \sum_{i=0}^{N}
\sum_{ j=i+1}^N
 \frac{k^2\,m_i\,m_j}{\Delta_{ij}} ,
\end{equation}
where the mutual distance $\Delta_{ij}$ is
\begin{equation}\label{Delta}
    \Delta_{ij} = ||\vec{p}_j - \vec{p}_i||,
\end{equation}
and $k$ stands for the Gaussian gravity constant.

The Hamiltonian \Eqref{Hbar} admits six integrals of barycenter
\begin{equation} \label{intbar}
   \sum_{i=0}^N m_i\,\vec{p}_i = \vec{0},  \qquad
  \sum_{i=0}^N \vec{P}_i  = \vec{0},
\end{equation}
as well as the energy integral $\cH = \mbox{const}$, and
three angular momentum integrals
\begin{equation}\label{Gint}
    \vec{G} = \sum_{i=0}^N \vec{p}_i \times \vec{P}_i = \mathbf{const}.
\end{equation}
The integrals are usually exploited as the accuracy control tool
when the differential equations of motion
\begin{equation}\label{Beq}
   \dot{ \vec{p}} = \frac{\partial \cH}{\partial \vec{P}},
    \qquad  \dot{\vec{P}} = - \frac{\partial \cH}{\partial \vec{p}},
\end{equation}
are solved numerically.

Instead of solving the $6N+6$ order system \Eqref{Beq}, it is often desirable to
study the relative motion of $N$ bodies with respect to the material point
$m_0$. But if the integration method to be applied is symplectic, it is
necessary to use the Hamiltonian equations of motion, hence the necessity of
defining a canonical transformation $(\vec{p},\vec{P},\cH) \leftrightarrows
(\vec{r},\vec{R},\cK)$ with new, relative coordinates $\vec{r}$ and momenta
$\vec{R}$.

A naive, straightforward approach would consist in postulating $ \vec{r}_i =
\vec{p}_i - \vec{p}_0$ for all $i$. This leads to $\vec{r}_0 = \vec{0}$ and the
Jacobian matrix of the transformation becomes singular; such a transformation
cannot be canonical regardless of the choice of the momenta. However, the
difficulty can be easily circumvented if we change the definition of
$\vec{r}_0$. \citet{Poincare:96} proposed
\begin{equation}
\label{pt}
\begin{array}{rcl}
  \vec{r}_i &=& \vec{p}_i - \vec{p}_0, \qquad \mathrm{for~}i=1,\ldots,N, \\
  \vec{r}_0 &=& \vec{p}_0, \\
\end{array}
\end{equation}
whereas \citet{DLL:98} chose $\vec{r}_0$ as the position of the barycenter in
some arbitrary inertial frame. For the barycentric system, the latter choice
amounts to a ''differentiable zero'' $\vec{r}_0 = \left( \sum_{i=0}^N m_i
\right)^{-1} \sum_{i=0}^N m_i \vec{p}_i = \vec{0}$. It turns out, that both
starting points lead to the same result, so we continue our presentation
assuming the Poincar\'e choice \Eqref{pt}.

Retaining $\vec{r}_0$ as the position of the reference body $m_0$ with respect
to the barycenter, we can perform a canonical extension of the time-independent
point transformation \Eqref{pt},
{ 
requesting the
Mathieu transformation condition
}
\citep[p.~301]{Whittaker}
\begin{equation}\label{canext}
    \vec{P} \cdot  \mathrm{d}\vec{p} =  \vec{R} \cdot  \mathrm{d}\vec{r},
\end{equation}
or, explicitly,
\begin{equation}\label{canext1}
\sum_{i=0}   \vec{P}_i \cdot  \mathrm{d}\vec{p}_i =
 \vec{R}_0 \cdot  \mathrm{d}\vec{p}_0
 + \sum_{i=1}^N \vec{R}_i \cdot
 \left( \mathrm{d}\vec{p}_i - \mathrm{d}\vec{p}_0\right) .
\end{equation}
Equating the  coefficient
of each differential $\mathrm{d}\vec{p}_i$ to zero, we find the
new momenta
\begin{equation}
\label{ptm}
\begin{array}{rcl}
  \vec{R}_i &=& \vec{P}_i, \qquad \mbox{for} \qquad i=1,\ldots,N, \\
  \vec{R}_0 &=&  \sum_{i=0}^N \vec{P}_i ~=~\vec{0}. \\
\end{array}
\end{equation}
The fact, that $\vec{R}_0=\vec{0}$ is a direct consequence of the
integrals of barycenter \Eqref{intbar}.

Equations \Eqref{pt} and \Eqref{ptm}, that are due to \citet{Poincare:96},
define the canonical relative variables. The coordinates $\vec{r}$ consist of
the positions with respect to the reference body $m_0$, save for the $\vec{r}_0$
that is measured with respect to the barycenter. The momenta $\vec{R}$ are
measured with respect to the barycenter, save for $\vec{R}_0$ that can be
understood as measured with respect to $m_0$ and hence it is zero (although with
nonvanishing partials with respect to $\vec{P}_i$).

The most important feature is that the new Hamiltonian $\cK$, obtained by the
simple substitution of the transformation equations into $\cH$, does not depend
neither on the coordinates, nor on the momenta of $m_0$. Indeed, one obtains
\citep{Poincare:96,Whittaker,Hagihara:70,Deprit:83,Laskar:90,Laskar:91}
\begin{eqnarray}
  \cK & = & \frac{1}{2} \sum_{i=1}^{N}  \left( \frac{1}{m_0}
  + \frac{1}{m_i} \right)
  \vec{R}_i^2   + \frac{1}{m_0}
  \sum_{i=1}^{N}  \sum_{j=i+1}^{N}
\vec{R}_i \cdot \vec{R}_j  - \nonumber \\
& & -
 \sum_{i=1}^{N} \frac{k^2 m_0 m_i}{r_i} -
 \sum_{i=1}^{N} \sum_{j=i+1}^{N} \frac{k^2 m_i m_j}{\Delta_{ij}}, \label{Nc:hamP}
\end{eqnarray}
where $\Delta_{ij}=||\vec{p}_j-\vec{p}_i|| = ||\vec{r}_j-\vec{r}_i||$,
or, equivalently \citep{DLL:98,Chambers:99}
\begin{eqnarray}
  \cK & = & \frac{1}{2} \sum_{i=1}^{N}
  \frac{1}{m_i}
  \vec{R}_i^2   + \frac{1}{2m_0}\left(
  \sum_{i=1}^{N}
\vec{R}_i  \right)^2 - \nonumber \\
& & -
 \sum_{i=1}^{N} \frac{k^2 m_0 m_i}{r_i} -
 \sum_{i=1}^{N} \sum_{j=i+1}^{N} \frac{k^2 m_i m_j}{\Delta_{ij}}. \label{HDLL}
\end{eqnarray}

An important fact to be remembered is that the Hamiltonian $\cK$ has the form
\Eqref{Nc:hamP} or \Eqref{HDLL} only if the substitution of the barycenter 
integrals \Eqref{intbar} has been performed. Thus it cannot serve to obtain the
equations for $\dot{\vec{r}}_0$ or $\dot{\vec{R}}_0$. However, once we know all
the remaining $\vec{r}_i$ and $\vec{R}_i$, the $\vec{r}_0$ values can be  easily
computed from Equations \Eqref{intbar} and \Eqref{pt}, whereas
$\vec{R}_0=\vec{0}$ by the definition. For all the remaining bodies
\begin{equation}\label{Peq}
    \dot{\vec{r}}_i = \frac{\partial \cK}{\partial \vec{R}_i},
    \qquad  \dot{\vec{R}}_i = - \frac{\partial \cK}{\partial \vec{r}_i},
\end{equation}
and we will assume that $i=1,\ldots,N$ throughout the rest of this
paper.

A remarkable property of the Poincar\'e variables is
\begin{equation}\label{GP}
    \vec{G} = \sum_{i=0}^N \vec{p}_i \times \vec{P}_i =
    \sum_{i=1}^N \vec{r}_i \times \vec{R}_i,
\end{equation}
which means, that the total angular momentum of the reduced
system of $N$ bodies evaluated by means of the
Poincar\'e variables is the same as the angular momentum
of $N+1$ bodies evaluated in the barycentric frame.

If the reference body mass $m_0 \gg m_i$, the Hamiltonian $\cK$ can be easily
partitioned into the unperturbed, Keplerian part and a small perturbation
proportional to the greatest of $m_i$. From the point of view of analytical
theories  using these variables \citep{YuasaHori:79,Hori:85}, it is preferable
to split $\cK$ into the sum
\begin{equation}\label{parta}
    \cK = \cK_0^{(a)} + \cK_1^{(a)},
\end{equation}
where \citep{Laskar:90,Laskar:91}
\begin{eqnarray}
  \cK_0^{(a)} & = & \frac{1}{2} \sum_{i=1}^{N}
   \frac{m_0+m_i}{m_0\,m_i} \vec{R}_i^2    -
 \sum_{i=1}^{N} \frac{k^2 m_0 m_i}{r_i},
 \label{Nc:K0}\\
 \cK_1^{(a)} & = &   \frac{1}{m_0}
  \sum_{i=1}^{N}  \sum_{j=i+1}^{N}
\vec{R}_i \cdot \vec{R}_j  -
 \sum_{i=1}^{N} \sum_{j=i+1}^{N} \frac{k^2 m_i m_j}{\Delta_{ij}} .
 \label{Nc:K1}
\end{eqnarray}
The principal part $\cK_0^{(a)}$ defines $N$ \emph{relative} two-body problems.
The perturbation $\cK_1^{(a)}$ involves not only the mutual interactions between
the minor bodies $m_i$, but also the momenta related terms that replace the
usual ''indirect part'' of the perturbing function present in noncanonical
relative $(N+1)$-body problem \citep{Poincare:05}. It is due to this term, that
the Poincar\'e variables were considered somehow handicapped; the objection that
velocities are no longer tangent to the momenta became almost a proverb,
although many non-inertial reference frames have the same property, the
restricted three-body problem being the best example. This objection has
fortunately ceased to be taken seriously; for example, \citet{FMM:04}
successfully use orbital elements evaluated from the Poincar\'e momenta
$\vec{R}$. In this paper we will use alternatively two types of orbital
elements. The \textit{osculating elements} are computed by the usual two body
formulae from astrocentric  positions $\vec{r}_i$ and velocities
$\dot{\vec{r}_i}$; we use them in all plot labels and orbital data tables. But
in the calculation of spectral numbers or in the definitions of resonance
arguments, we use orbital elements computed from $\vec{r}_i$ and
$m_0m_i/(m_0+m_i)\,\vec{R}_i$, calling them \textit{contact elements} after
\citet{Brumberg91}. 
[{In fact, the transformation between astrocentric positions -- barcentric momenta
and the contact elements can be still done with the usual two-body formulae
\citep{Morbidelli:book}}]. The former are commonly used in literature, whereas
the latter offer a better behavior from the dynamical point of view. 
{
The superiority of contact elements results from the fact, that the reference frame
of Cartesian momenta is inertial, hence the influence of noninertial forces is
reduced to purely kinematical contribution. The inferred Keplerian angles
($\omega,\Omega, \mathcal{M}$) and the conjugate momenta can be interpreted as
canonical Delaunay's elements \citep{Morbidelli:book}, so  the derivation and
interpreation of the fundamental frequencies  is straightforward.
}

\subsection{Symplectic integration: two is a company}

With the advent of symplectic integrators based on the Wisdom-Holman approach
\citep{WH:91}, the Poincar\'e variables became an attractive framework for the
numerical studies of planetary systems \citep{DLL:98,Chambers:99}. First of all,
similarly to the Jacobian coordinates, they reduce 
{the number of
equations of motion} by 6 with respect to the barycentric problem. Thanks to the
possibility of splitting $\cK$ into the main part and a perturbation, they also
allow the construction of a symplectic integrator with the local truncation
error proportional to the product of $m_i/m_0$, hence a larger integration step
$h$ can be applied.

Given a Hamiltonian $\mathcal{M} = \mathcal{M}_0 + \varepsilon \mathcal{M}_1$
with a small parameter $\varepsilon$, the W-H integrator is based on the
alternating application of maps $\Phi_{0,\tau}(\vec{r},\vec{R})$ and
$\Phi_{1,\tau}(\vec{r},\vec{R})$ that represent the solutions of equations of
motion derived from $\mathcal{M}_0$ and $\varepsilon \mathcal{M}_1$ alone, on
the interval  from $t_0$ to $t_0+\tau $. 
{ 
Moreover, each of the Hamiltonian parts should admit (a possibly simple)
analytical solution of the equations of motion.
}

In the numerical applications, Hamiltonian $\cK$ is typically split differently
than in Equation \Eqref{parta}, namely \citep{DLL:98,Chambers:99}
\begin{equation}\label{partnew}
    \cK = \cK_0  + \cK_1,
\end{equation}
where
\begin{eqnarray}
  \cK_0 & = & \frac{1}{2} \sum_{i=1}^{N}
   \frac{1}{m_i} \vec{R}_i^2    -
 \sum_{i=1}^{N} \frac{k^2 m_0 m_i}{r_i},
 \label{K0n}\\
 \cK_1 & = &  \frac{1}{2 m_0} \left( \sum_{i=1}^{N}
    \vec{R}_i\right)^2    -
 \sum_{i=1}^{N} \sum_{j=i+1}^{N} \frac{k^2 m_i m_j}{\Delta_{ij}} .
 \label{K1n}
\end{eqnarray}
The unperturbed part $\cK_0$ has now a different meaning: it
still leads to $N$ relative two-body problems
\begin{eqnarray}\label{kep:1}
    \dot{\vec{r}}_i & = &  \frac{\partial \cK_0}{\partial \vec{R}_i} ~=~
    \frac{\vec{R}_i}{m_i}, \\
    \dot{\vec{R}}_i & = & - \frac{\partial \cK_0}{\partial \vec{r}_i}
     ~=~ - \frac{k^2 m_0
    m_i}{r_i^3}\,\vec{r}_i, \label{kep:2}
\end{eqnarray}
but this time they are the \emph{restricted} two-body problems
{ with negligible masses $m_i$ or a fixed center of gravity:}
\begin{equation}\label{kep:12}
    \vec{\ddot{r}}_i = - \frac{k^2 m_0}{r_i^3}\,\vec{r}_i.
\end{equation}

The perturbation part $\cK_1$ remains proportional to $m_i/m_0$
and it is still a function of both coordinates $\vec{r}$ and momenta
$\vec{R}$. \citet{DLL:98} and then \citet{Chambers:99} considered it an
obstacle, so they further split $\cK_1$ into
\begin{equation}\label{K1sp}
    \cK_1 = \cK_{11}(\vec{R}) + \cK_{12}(\vec{r}),
\end{equation}
obtaining elementary ''kick'' maps $\Phi_{11,\tau}$
\begin{eqnarray}
 \vec{r}'_i & = & \vec{r}_i  +  \frac{\tau}{m_0} \sum_{j=1}^{N} \vec{R}_j, \\
 \vec{R}'_i & = & \vec{R}_i .
\end{eqnarray}
and $\Phi_{12,\tau}$
\begin{eqnarray}
 \vec{r}'_i & = & \vec{r}_i , \\
 \vec{R}'_i & = &
  \vec{R}_i - \tau \,\sum_{j=1, j\neq i}^{N}
    \frac{k^2 m_i m_j}{\Delta_{ij}^3} \, \left( \vec{r}_i - \vec{r}_j \right).
\end{eqnarray}
In the formulas of both maps we add a prime to the symbols standing for the values of
coordinates and momenta at $t_0+\tau$, whereas unprimed symbols refer to the values at $t_0$.

As the effect of the partition \Eqref{K1sp}, the classical ''leapfrog''
\begin{equation}
\label{leapr}
  \Phi_\tau \approx  \Phi_{1,\tau/2} \circ \Phi_{0,\tau} \circ \Phi_{1,\tau/2},
\end{equation}
was replaced by
\begin{equation}
\label{leapinc}
  \Phi_\tau \approx  \Phi_{11,\tau/2} \circ \Phi_{12,\tau/2} \circ \Phi_{0,\tau} \circ \Phi_{12,\tau/2}
  \circ \Phi_{11,\tau/2}.
\end{equation}
According to \citet{DLL:98}, the ordering of $\Phi_{11}$ and $\Phi_{12,\tau}$ is
insignificant,
and, indeed,  \citet{Chambers:99} interchanged them, using
\begin{equation}
\label{leapch}
  \Phi_\tau \approx  \Phi_{12,\tau/2} \circ \Phi_{11,\tau/2} \circ \Phi_{0,\tau} \circ \Phi_{11,\tau/2}
  \circ \Phi_{12,\tau/2}.
\end{equation}
The interchange of the maps is justified by the fact that $\cK_{11}$ and $\cK_{12}$ commute, i.e.
the Poisson bracket $\{\cK_{11};\,\cK_{12}\}=0.$
In these circumstances
\begin{equation}
\label{compo}
  \Phi_{1,\tau} =   \Phi_{12,\tau} \circ \Phi_{11,\tau} =  \Phi_{11,\tau }
  \circ \Phi_{12,\tau},
\end{equation}
and we can concatenate both maps obtaining $\Phi_{1,\tau}$ in a compact form.
\begin{eqnarray}
\label{fi1a}
  \vec{r}'_i  &=& \vec{r}_i +
  \frac{\tau}{m_0} \sum_{j=1}^{N} \vec{R}_j \\
\label{fi1b}
\vec{R}'_i  &=& \vec{R}_i - k^2\, m_i
\,\tau\,\sum_{j=1, j\neq i}^{N}
    \frac{m_j \left( \vec{r}_i - \vec{r}_j
    \right)}{\Delta_{ij}^3},
\end{eqnarray}
where $\vec{r}_i$ stands for $\vec{r}_i(t_0)$,
$\vec{r}'_i$ stands for $\vec{r}_i(t_0+\tau)$, and similarly for
$\vec{R}_i$, $\vec{R}'_i$.

Separating $\cK$ according to Equations \Eqref{partnew}, \Eqref{K0n}, and
\Eqref{K1n} offers a possibility of using only two maps for a W-H integrator.
{ 
Although our partitioning seems to be more compact and clear
than the ``heliocentric-democratic'' scheme, we
note that both mappings are practically equivalent as far as the CPU cost
is concerned. Yet
the rule ``two is a company, three is a crowd''
} 
holds true in the realm of symplectic integrators for perturbed systems:
according to the theorem of \citet{Suzuki:91} any symplectic composition method
of an order higher than 2 must necessarily involve stages with negative
sub-steps that amplify accumulation of roundoff errors. This holds true for a
composition of maps derived from splitting the Hamiltonian into any number of
terms. However, \citet{LasRob} found a family of methods designed for two-terms
perturbed systems with $\cK = \mathcal{A} + \varepsilon \mathcal{B}$ where the
negative sub-steps are avoided. Their integrators do not contradict the results
of Suzuki: formally they remain second order methods, but in contrast to other
W-H methods with local truncation errors $O(\varepsilon \tau^2)$, their errors
have a form $O(\varepsilon^2 \tau^2 + \varepsilon \tau^n)$. So, for sufficiently
small perturbation $\varepsilon$, the Laskar-Robutel methods may behave like
higher order integrators in certain domain of stepsize $\tau$ although no
negative substeps were introduced.

\subsection{Tangent maps}

Whenever a differential correction of initial conditions or the computation of
sensitivity indicators is required, the use of tangent maps becomes
indispensable. Keplerian map $\Phi_0$ and its associate tangent map can be
computed according to a comprehensive recipe by \citet{MikIn:99}. The
propagation of a tangent vector
\begin{equation}\label{tvdef}
    \vec{\xi} = \left(%
\begin{array}{c}
  \delta\vec{r} \\
  \delta\vec{R} \\
\end{array}%
\right),
\end{equation}
under the action of the ``kick'' map $\Phi_1$ amounts to multiplying it
by the Jacobian matrix $\mathrm{D}\Phi_1$. Resulting expressions are simple:
\begin{eqnarray}\label{tv1}
    \delta\vec{r}' & =& \delta\vec{r} + \frac{\tau}{m_0} \sum_{j=1}^N \delta\vec{R}_j, \\
    \delta\vec{R}_i' & =& \delta\vec{R}_i +  \tau \,k^2\,m_i \sum_{j=1,\,j \neq i}^N
    \frac{m_j}{\Delta_{ij}^3} \left[ \vec{\delta}_{ij}
    + \frac{3\,\delta\Delta_{ij}}{\Delta_{ij}}\,(\vec{r}_i-\vec{r}_j) \right],
\end{eqnarray}
where
\begin{equation}\label{dd}
    \delta\Delta_{ij} = \frac{\left(\vec{r}_j-\vec{r}_i \right) \cdot \vec{\delta}_{ij}}{\Delta_{ij}} ,
    \qquad
    \vec{\delta}_{ij} =  \delta\vec{r}_j-\delta\vec{r}_i .
\end{equation}

\subsection{Angular momentum integral}

It can be easily demonstrated, that any composition of maps $\Phi_0$ and $\Phi_1$ conserves
the angular momentum integral (\ref{GP}). This property is guaranteed by the conservation of $\vec{G}$
by each map separately. Recalling that $\Phi_0$ and $\Phi_1$ define \emph{exact} solutions
of motion generated by $\cK_0$ and $\cK_1$ respectively, we can simply check that
\begin{equation} \label{Gcon}
 \{\vec{G};\,\cK_0\} = \{\vec{G};\,\cK_1\} = \vec{0}.
\end{equation}
The proof of Eq.~(\ref{Gcon}) is straightforward. Starting from the definition of
$\vec{G}$, we use the linearity of Poisson brackets and the Leibnitz identity to write
\begin{eqnarray}
  \{\vec{G};\,\cK_0\}  & = & \sum_{i=1}^N \vec{r}_i \times \{\vec{R}_i;\,\cK_0\}
  -  \vec{R}_i \times \{\vec{r}_i;\,\cK_0\} = \nonumber \\
  &=& - \sum_{i=1}^N \vec{r}_i \times \frac{\partial\,\cK_0}{\partial \vec{r}_i}
  +  \vec{R}_i \times \frac{\partial\,\cK_0}{\partial \vec{R}_i}.
\end{eqnarray}
Then we substitute the right-hand sides of Equations (\ref{kep:1}) and (\ref{kep:2}), concluding that
all vector products vanish and indeed $ \{\vec{G};\,\cK_0\} = \vec{0}$. A similar procedure demonstrates
$ \{\vec{G};\,\cK_1\} = \vec{0}$.

Of course, from  practical point of view the conservation of $\vec{G}$ is only up to
computer roundoff errors.

\subsection{Chaoticity indicators}
\label{S:MEGNO}

To detect unstable motions in the phase space, many numerical tools are
available. Concerning the dynamics of close-to-integrable Hamiltonian systems,
they can be roughly divided in two classes: spectral algorithms that resolve the
fundamental frequencies and/or their diffusion rates
\citep{Laskar1993,Nesvorny1997,Michtchenko2001}, and methods based on the
divergence rate of initially close phase trajectories, expressed in terms of the
Lyapunov exponents \citep{Benettin1980,Froeschle1984}.

In this work, among the the spectral tools, we choose the method invented by
\citet{Michtchenko2001}; its idea is genuinely simple --- to detect chaotic
behavior one counts the number of frequencies in the FFT-spectrum of an
appropriately chosen dynamical signal. We deal with conservative Hamiltonian
systems; so in a regular case, the spectrum of fundamental frequencies is
discrete and we obtain only a few dominant peaks in the FFT spectrum. Chaotic
signals do not have well defined frequencies, and their FFT spectrum is very
complex. The number of peaks in the spectrum above some noise level $p$
(typically, $p$ is set to a few percent of the dominant amplitude) tell us on
the character (regular, chaotic) of the of the system.

{ 
The method by \citet{Michtchenko2001} does not have as strong theoretical
foundations as the Frequency Map Analysis (FMA) by \cite{Laskar1993} or the
Fourier Modified Transform (FMT) by \cite{Nesvorny1997} which are considered as
rigorous and efficient tools. We did some comparative tests of the later
algorithm with MEGNO already \citep{Gozdziewski2004}. Here, we choose the
method of \cite{Michtchenko2001} for its appealing simplicity and because we
used it in the former papers devoted to the analysis of the RV data. In that
way, we can compare the results directly. In our code, the spectral signals
analyzed with the FFT are related to canonical Poincar\'e elements, so the
fundamental frequencies are well defined. Moreover, we resolve the chaotic and
regular signals by comparing the number of significant peaks in the FFT
spectrum, thus a very precise determination of the fundamental frequencies is
not critical. Actually, in this work we also show that the algorithm has same
drawbacks and should be applied with care.
}

The basic tool to discover exponentially unstable bounded orbits, i.e. chaotic
orbits, is the Maximum  Lyapunov Characteristic Exponent (MLCE) $\sigma$.
Numerical symplectic integration methods are fixed step algorithms, so we can
restrict our discussion to the iterations of a discrete map $\vec{\zeta}_n =
\Phi^n \vec{\zeta}_0$, that generates a sequence of state vectors
$\vec{\zeta}_n$ consisting of coordinates and their conjugate momenta. The
direct computation of the MLCE is based on the analysis of the tangent vectors
$\vec{\delta}_n$ that evolve under the action of a linear tangent map
$\vec{\delta}_n = (D\Phi)^n \vec{\delta}_0$. Asymptotically, the MLCE value is
given by
\begin{equation} \label{sig}
    \sigma = \lim_{n \rightarrow \infty} \frac{1}{n} \sum_{k=1}^n
    \ln \left(\frac{\delta_k}{\delta_{k-1}}\right).
\end{equation}
If $\sigma$ converges to some positive value, we conclude that the nominal orbit
$\vec{\zeta}_n$ and some initially close orbit diverge exponentially at the rate
$\exp(\sigma t)$. Two practical difficulties arise when the direct definition
(\ref{sig}) is used: the convergence of $\sigma$ is often very slow, and it is
difficult to tell how small should be the final value of $\sigma$ to consider it
$\sigma=0$.

A large variety of methods has been proposed to overcome the problem of slowly
convergent MLCE estimates. The authors prefer the so called MEGNO (Mean
Exponential Growth factor of Nearby Orbits) indicator proposed by
\citet{Cincotta2000} -- that choice is justified by the successful application
of this method in our previous works \citep[e.g.,][and the references
therein]{BMBW:2005,Gozdziewski2006a}. The definition of MEGNO  for a discrete
map is \citep{Cincotta2003}
\begin{equation}\label{mfm2}
    Y(n) = \frac{1}{n} \sum_{k=1}^n  y(k),
\end{equation}
where
\begin{equation}\label{mfm1}
    y(n) = \frac{2}{n} \sum_{k=1}^n k\,\ln\left(
    \frac{\delta_k}{\delta_{k-1}}\right).
\end{equation}

If the iterates of the discrete map refer to the moments of time
separated by the stepsize $h$, the discrete map MEGNO function $Y(n)$ asymptotically
tends to
\[
  Y_n = a\,h\,n+b,
\]
with $a=0, b=2$ for a quasi-periodic orbit, $a=b=0$ for a stable, isochronous
periodic orbit, and $a = \frac{1}{2} \sigma,\,b=0$ for a chaotic orbit. Thus we
can indirectly estimate the MLCE on a finite time interval, but the weight
function $k$ in the sum (\ref{mfm1}) reduces the contribution of the initial
part of the tangent vector evolution, when the exponential divergence is to
small to be observed behind other linear and nonlinear effects
\citep{Morbidelli:book}. Thus, fitting the straight line to the final part of
$Y(n)$, we obtain good estimates of $\sigma$ from a relatively shorter piece of
trajectory than in the direct MLCE evaluation.

In practical application, one can use a more convenient form of
Eqs.~(\ref{mfm2}) and (\ref{mfm1}) proposed by \cite{BMBW:2005}
\begin{eqnarray} \label{updY}
        Y(n)  & = &  \frac{(n-1) \, Y(n-1) + y(n)}{n}, \\
\label{updy}
        y(n) & =  &
         \frac{n-1}{n}\,y(n-1) + 2 \, \ln \left(\frac{\delta_n}{\delta_{n-1}}\right),
\end{eqnarray}
with the initial setup $y(0)=Y(0)=0$. The fact that only the ratio
$\delta_n/\delta_{n-1}$ is significant, as well as the linearity of tangent map,
allows to avoid the overflow of $\delta_n$ thanks to occasional normalization of
the tangent vector length to $\delta_n=1$ performed after the ratio of
$\delta_n/\delta_{n-1}$ has been evaluated.

\section{Stability of the HD 37124 planetary system}

As a non-trivial application of the presented algorithms and the illustration of
difficulties arising in the dynamical analysis of the long-term stability of
multiplanet configurations, we choose the HD~37124 extrasolar system. The
discovery of two Jovian planets has been announced by \cite{Butler2001} and
confirmed by \cite{Vogt2005}. At first, the system seemed to be well modeled by
a 2-planet configuration \citep{Butler2001}. However, new observations lead to
two-planet fits with $e_c \sim 0.7$ and a catastrophically unstable
configuration. Moreover, with the updated RV observations,  \citet{Vogt2005}
found much better model of 3 planets with similar masses of
$\sim0.6$~$m_{\idm{J}}$ in low-eccentric orbits.  The best fits have the rms
$\sim4$~m/s, in agreement with the internal accuracy of the data. However, the
best-fit orbital solution, both in the the kinematic Keplerian model, and in
more realistic $N$-body simulation (see Table~1), lies close to the collision
line of planets c~and d. Note, that  we define the collision line in terms of
semi-axes and eccentricities as $a_c (1+e_c) = a_d (1-e_d)$.  This line marks
the zone in which the mutual interactions of massive companions can quickly
destabilize the system.

\begin{table}
\label{tab:tab1}
\caption{
The bet-fit astro-centric, osculating Keplerian elements of a stable  HD~37124
planetary configuration at the epoch of the first observation
$t_0$=JD2,451,0420.047. Mass of the parent star is 0.78~$m_{\sun}$. The fit has
been refined with GAMP over  $\sim 5\cdot 10^4 P_{\idm{d}}$. See
\citep{Vogt2005,Gozdziewski2006a} and Fig.~\ref{fig:fig2} for more details.
}
\centering
\begin{tabular}{lccc}
\hline
Parameter \hspace{1em}
& \ \ planet b \ \  & \ \ planet   c  \ \ & \ \ planet   d
\\
\hline
$m \sin i$ [m$_{\idm{J}}$]
                 &   0.62447   &    0.56760    &   0.71194
\\
$a$ [AU]         &   0.51866   &   1.61117    &   3.14451
\\
$e$              &   0.07932   &    0.15267    &   0.29775
\\
$\omega$ [deg]   &  138.405    &  268.863      & 269.494
\\
${\cal M}(t_0)$ [deg]
                 &  259.011    &  109.545      & 124.113
\\
$\Chi$         & \multicolumn{3}{c}{0.938}
\\
$V_0$ [m s$^{-1}$]         & \multicolumn{3}{c}{7.629}
\\
rms~[m s$^{-1}$]       & \multicolumn{3}{c}{3.39}
\\
\hline
\end{tabular}
\end{table}


How to interpret the RV measurements remains an open question. The dynamical
long-term stability of the planetary system is the most natural requirement of a
configuration consistent with observations. Yet the three-planet model is
parameterized by at least 16~parameters, even assuming that the system is
coplanar. For that reason the search for the best fits  fulfilling the
constraints of stability is a difficult task. It can be resolved in different
ways. For example, we may try to find dynamically stable solutions in the
vicinity of the formal best fit configurations (the latter are often unstable).
However, examining the stability of configurations in that neighborhood, we have
no reasons to expect that the stable fits are optimal in the statistical sense.
Another approach relies on the elimination of unstable fits {\em during} the
fitting process, through {\em penalizing} unstable solutions with a large value
of $\Chi$. This method, described in  \citet{Gozdziewski2006a}, is dubbed GAMP
(Genetic Algorithm with MEGNO Penalty). It was shown, that such an approach is
particularly useful in modeling resonant or close-to-resonant planetary
configurations.
{ 
Unfortunately, the  algorithm cannot give definite answer when we want to
resolve the $\Chi$ shape of strictly stable solutions in detail. The penalty
term in the $\Chi$ function relies on a signature of the system stability,
expressed through the fast indicator. Due to significant CPU overhead, the fast
indicator in the minimizing code can be only calculated over relatively short
time, typically $10^3$ orbital periods of the outermost planet. Moreover, the
code can converge to unstable best fits that appear stable on that short time
scale. Hence, at the end of the search, we have to examine the stability of the
individual best fits in the obtained ensemble of solutions,  over the time-scale
of relevant mean motion and secular resonances.
}

\begin{figure*}
\centerline{
  \includegraphics[width=5.2in]{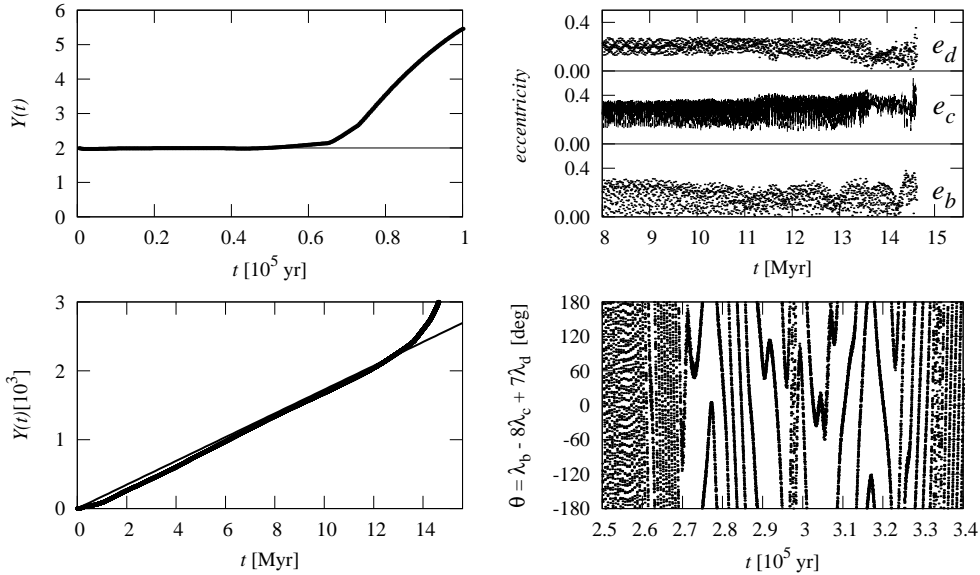}
}
\caption{
Evolution of the HD~37124 system selected  best fit related to the  theree-body
MMR  (initial conditions listed in the text). Left: evolution of MEGNO over a
short (top) and a long (bottom) time. The straight line is the least-square fit
to $Y(t)=(\sigma/2) t +b $. Top right: contact eccentricities during
$\sim15$~Myr. Bottom right: the critical argument of the three-body MMR
$+1\mbox{b}:-8\mbox{c}:+7\mbox{d}$.
}
\label{fig:fig1}
\end{figure*}

\subsection{Long-term stability of the best-fit configurations}

\begin{figure*}
\includegraphics[]{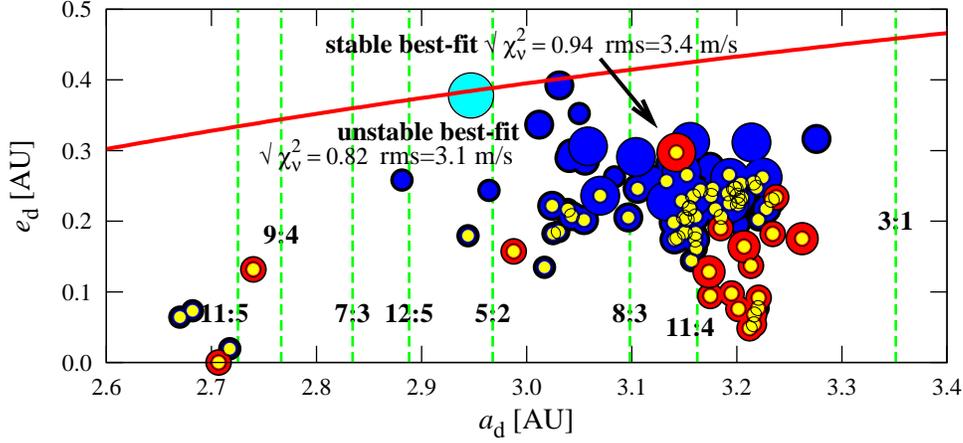}
\caption{
The long-term stability of the best-fits obtained in the GAMP search by
\citet{Gozdziewski2006a}. The osculating elements at the epoch of the first
observation (JD~2,450,420.047) are projected onto the ($a_d,e_d$)-plane (i.e.,
the semi-major axis vs the eccentricity of the outermost planet). The quality of
fits in terms of $\Chi<1.1$ and rms$<4$~m/s is marked by the size of blue (dark
grey) and red (medium grey) circles (larger circles have better fit quality).
The best fit, self-consistent Newtonian
configuration obtained without stability constraints, in terms of  quintuples
($m$~[m$_{\idm{J}}$], $a$~[AU], $e$, $\omega$~[deg], ${\cal M}$~[deg]) at the
epoch of the first observation, is
(0.619,   0.519,   0.088, 141.91, 257.34),
(0.565,   1.663,   0.104, 331.88, 67.71), and
(0.732,   2.947,   0.378, 283.33, 95.32)
for planets b, c, and d, respectively, with velocity offset $7.53$~m/s. The blue
circles are for chaotic solutions that did not survive the integration time of
$\sim37$~Myr. The red circles are for regular solutions --- in that case the
MEGNO converged to~2. The small yellow (light grey) circles mark configurations
that survived the integration.  The best {\em stable} fit found is marked with
an arrow, its osculating elements are given in Table~1.  Some dominant MMRs of
planets c and d are marked with dashed vertical lines, according to the third
Kepler law, and labeled accordingly. The red curve marks the collision line of
the two outermost orbits.
}
\label{fig:fig2}
\end{figure*}

The GAMP analysis of the RV data of HD~37124 published by \citet{Vogt2005} was
presented in \cite{Gozdziewski2006a}.  About 100 of best fit solutions were
found yielding $\Chi<1.1$, the rms $\sim4$~m/s, and {\em stable} in the sense
that their MEGNO signatures are close to 2 up to $\sim1000-2000$ orbital periods
of the outermost planet. Due to heavy CPU requirements,  the time-span to
resolve MEGNO in the GAMP code cannot be set very long. The short integration
time $\sim10^3 P_{\idm{d}}$ allows only to eliminate strongly chaotic
configurations, typically leading to collisions between  planets and/or with the
parent star. The best fit solutions were found in a dynamically active region of
the phase space, spanned by a number of low-order mean motion resonances (MMRs)
between the two outermost Jovian companions, like 5c:2d,  8c:3d, or 11c:4d (see
Fig.~\ref{fig:fig2} and dynamical maps presented in
Figs.~\ref{fig:fig3},\ref{fig:fig4},\ref{fig:fig5},\ref{fig:fig6}). In
particular, close to the collision lines, the low-order MMRs overlap, giving
rise to the global instability zone.

Yet we should be aware that two-body MMRs with characteristic time scale $\sim
10^4$--$10^5$ orbital periods of the outermost planet are not the only source of
instability in the multi-planet system. Already when we deal with three-planet
configurations, the strong instabilities may be generated by three-body MMRs or
by long-term secular resonances
\citep{Nesvorny1998,Nesvorny1999,Murray1998,Guzzo2006}.  In such instance,
appropriately longer integration time is necessary to detect the unstable
solutions.

This issue is illustrated in Fig.~\ref{fig:fig1}. We choose one of the best fits
with  initial osculating, astrocentric Keplerian elements at the epoch  of the first
observation in terms of tuples ($m$~[m$_{\idm{J}}$], $a$~[AU], $e$,
$\omega$~[deg], ${\cal M}$~[deg]):
(0.593, 0.519, 0.0058, 303.360, 95.060),
(0.558, 1.615, 0.101, 315.621, 70.279), and
(0.690, 3.193, 0.26111, 255.848, 142.886)
for planets b, c, and d, respectively. These initial conditions had
$\Chi \approx 0.98$ and an rms about $4$~m/s.
In the time range covered by the GAMP integrations (and up to $\sim10^4
P_{\idm{d}}$, i.e. $\sim60,000$~yr), the configuration appears  strictly regular
because the indicator quickly converges to~2 (the top-left panel in
Fig~\ref{fig:fig1}). Nevertheless, after the transient time, the MEGNO starts to
grow linearly at the rate of $\sigma/2 \sim2\cdot 10^{-4} \mbox{yr}^{-1}$ (where
$\sigma$ is the MLCE of the solution, see bottom-left panel in
Fig.~\ref{fig:fig1}). Actually, after a relatively long time $\sim15$~Myr, the
chaotic motion leads to a collision between planets c~and d (the top-right
panel) due to a sudden increase of both eccentricities up to 0.6. The
elimination of such solutions during an extensive GAMP-like search on a Myrs
interval would be very difficult.

Looking for the source of such dramatically unstable  behavior, we perform the
frequency analysis of the orbits with the MFT by \citet{Nesvorny1997}. Denoting
the proper  mean motions by $n_\mathrm{b},
n_\mathrm{c}$, and
$n_\mathrm{d}$, respectively, we found that
\[
 n_\mathrm{b} - 8 n_\mathrm{c} + 7 n_\mathrm{d} \approx -0.4^{\circ}/\mbox{yr},
\]
clearly indicating the three-body MMR of the first order, and we label it with
$+1\mbox{b}:-8\mbox{c}:+7\mbox{d}$. The time evolution of the critical argument
$\theta=\lambda_\mathrm{b} - 8 \lambda_\mathrm{c} + 7 \lambda_\mathrm{d}$ is
illustrated in the bottom-right panel of Fig.~\ref{fig:fig1}. The circulation of
the critical angle alternates with libration, indicating the separatrix
crossings that explain chaotic evolution.

The presented example has inspired us to follow a two-stage procedure in
modeling the RV data.  First, with a GAMP-like code we look for many best-fit
solutions, ideally, approximating the global shape of $\Chi$ and simultaneously
stable, at least over a relatively short period of time. At that stage the
stability constraints cannot be tight, not only  due to significant CPU
requirements but also because we should not discard weakly chaotic solutions.
Such configurations may be bounded over very long time, longer by orders of
magnitude than their Lyapunov time $T_{\idm{L}}=1/\sigma$. In the next step, we
either refine the search in a zone bounded by the previously found fits with
much longer integration times (still numerically expensive), or we examine each
fit with long-term direct integrations and/or evaluate a fast indicator
signature, like the MLCE, Spectral Number, or the diffusion of fundamental
frequencies.

Here, for each solution with $\Chi<1.1$, we computed its MEGNO signature. The
integration time span is  about of 37~Myr -- long enough to detect the relevant
chaotic three-body resonances and strong secular resonances. Here, and in the
experiments described later on, we use the SBAB3 integrator scheme by
\citet{LasRob}. The time-step is 4~days. The secular periods in the given range
of $a_\mathrm{d}$ are quite short, $\sim10^{4}$~yr, nevertheless we can expect that
dynamical effects of potentially active secular resonances could be detected
after thousands of such characteristic periods, hence counted in
$10^6$--$10^7$~yr. Figure~\ref{fig:fig2} illustrates the results. The quality of
fits in terms of $\Chi$ is marked by the size of circles (better fits have
larger circles). Red (medium grey) circles are for stable, quasi-periodic
solutions. In that case the system may be stable over a very long time. Blue
(dark grey) circles are for chaotic solutions that led to collisions between
planets and that did not survive during the integration time (the integrations
are interrupted if any of the eccentricities increases above 0.66). Finally,
small yellow (light grey) circles mark all configurations (not necessarily
regular) that survived, remaining bounded during the maximal integration time.
Clearly, most of  solutions with initial $e_\mathrm{d}>0.2$ are both chaotic and
unbounded. Nevertheless, some chaotic solutions appear on the borders of stable
regions as well. Generally, the distribution of fits gives us a clear image of
the border of global instability of the system, relatively far from the
collision zone.

\subsection{Fine structure of the phase space}

\begin{figure*}
\centerline{
  \includegraphics[width=14.6cm]{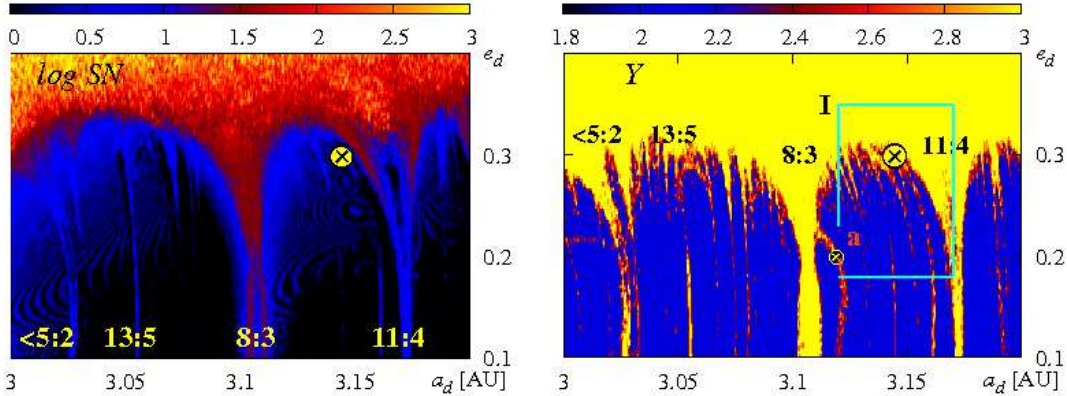}
}
\caption{
Dynamical maps of the Spectral Number (left) and the MEGNO indicator (right)
computed in the neighborhood of the best-fit solution to the RV data of HD~37124
with the integration time $\sim10^5~\mathrm{yr}$. The elements of the best fit
(see Table~1) are labeled with the crossed circle. The rectangle (I) marks the
borders of the close-up shown in Fig.~\ref{fig:fig4}. The stability of orbits is
color-coded: in both maps, yellow (pale grey) means strongly chaotic and
unstable solutions; regular configurations are marked black  in the SN-map and
dark blue (dark grey) in the MEGNO map.
}
\label{fig:fig3}
\end{figure*}

Figure~\ref{fig:fig3} compares the sensitivity of MEGNO and the Spec\-tral
Number when we use the same integration time, $\sim 10^5~\mathrm{yr} \approx 1.6
\cdot 10^4~P_{\idm{d}}$. In the case of the SN   map, we did the FFT on
$N=2^{19}$~steps of 64~d, counting the number of spectral lines  above $1\%$ of
the largest amplitude in the signal of  $f(t) = a_{\idm{c}}(t) \exp
\mbox{i}\lambda_{\idm{c}}(t)$, where $a_{\idm{c}}$ and $\lambda_{\idm{c}}$
denote the {\em contact} semi-major axis and mean longitude of planet~c.

Both dynamical maps  present the same region of the phase space, in the
neighborhood of the best fit. Note, that this particular solution has been
refined with GAMP integration over  time $5\sim10^4~P_{\idm{d}}$ that is about
of 2 orders of magnitude longer than in the set of selected solutions.  The
resolution of the maps is the same: $480 \times 120$ data points; the map
coordinates are usual astrocentric  osculating Keplerian elements. Most of
best-fits from Fig.~\ref{fig:fig2} lie in the region covered by
Fig.~\ref{fig:fig3}.

Both maps reveal a  number of unstable resonances. Yet the SN map involves some
artifacts (\emph{moire}-like patterns) related to a low value of the noise level
parameter $p$. Within the same integration time, the MEGNO map reveals
relatively more fine details than SN. In particular, the sophisticated border of
the collision zone appears to be more sharp and shifted towards smaller
$e_{\idm{d}}$. We can also  find some fine resonance lines entirely absent in
the SN map. For instance, there is a fine structure on the right-hand side of
the $8\mbox{c}:3\mbox{d}$~MMR. In order to investigate that instability, we
choose the initial condition marked with small crossed circle and labeled
with~\textbf{a}. The results of the MFT frequency analysis of this solution tell
us that the structure is related to the $+2\mbox{b}:-12\mbox{c}:+3\mbox{c}$ MMR.
The time evolution of the related critical argument is illustrated
Fig.~\ref{fig:fig8}a.

\begin{figure*}
\centerline{
  \includegraphics[width=14.6cm]{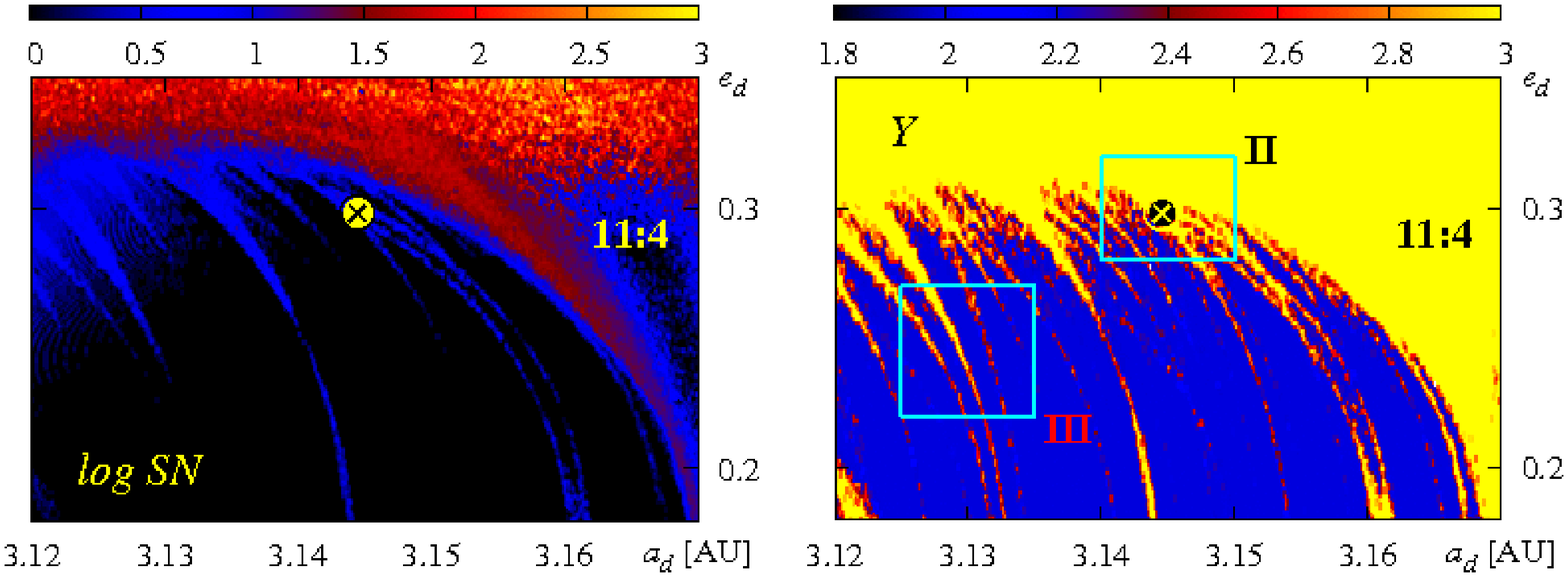}
 }
\caption{
Dynamical maps in terms of the Spectral Number (the left panel) and the MEGNO
indicator (the right panel)  computed in the region marked with small
rectangle~I in Fig.~\ref{fig:fig4}. The integration time $\sim3\times 10^5$~yr
is equivalent to $\sim4.8 \cdot 10^4$ orbital periods of the outermost planet.
The resolution of the maps $500\times120$ data points. The stability of orbits
is color-coded, see the caption to the previous figure.
}
\label{fig:fig4}
\end{figure*}

\begin{figure}
  \centerline{\includegraphics[width=7cm]{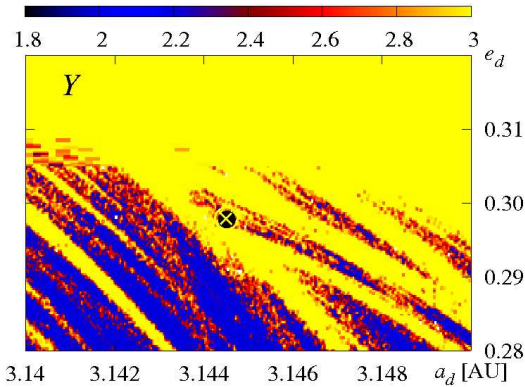}}
\caption{
Dynamical maps in terms of the MEGNO indicator  computed in the region marked by
small rectangle in Fig.~\ref{fig:fig4}. The integration time is $\sim3\times
10^5$~yr that is equivalent to $\sim4.8 \cdot 10^4$ orbital periods of the
putative outermost planet. The resolution of the plot is $200\times320 $ data
points.
}
\label{fig:fig5}
\end{figure}

Next, we computed close-ups of the dynamical map  within the rectangle labeled~I
in Fig.~\ref{fig:fig3}. These maps are shown in Fig.~\ref{fig:fig4}. This time
we increased $p$ to $5\%$ and the total number of steps has been doubled
($N=2^{20}$) in order to avoid the moire artifacts. But once again the
''concurrent'' MEGNO  map (the right panel of Fig.~\ref{fig:fig4}) calculated
over the same total time seems to offer a better representation of the phase
space. Interestingly, the best fit data (Table~1) seem to lie on the border of a
chaotic zone spanned by many overlapping resonances. A close-up of that area,
marked with rectangle~II in Fig.~\ref{fig:fig4}, is shown in
Fig.~\ref{fig:fig5}. Clearly, even a very small change of parameters of the
outermost planet may push the system into a strongly chaotic state.  It also
illustrates the good performance of the GAMP algorithm that was able to locate
and preserve the fit in an extremely narrow island of stable motions.

\begin{figure*}
\centerline{
  \includegraphics[width=14.6cm]{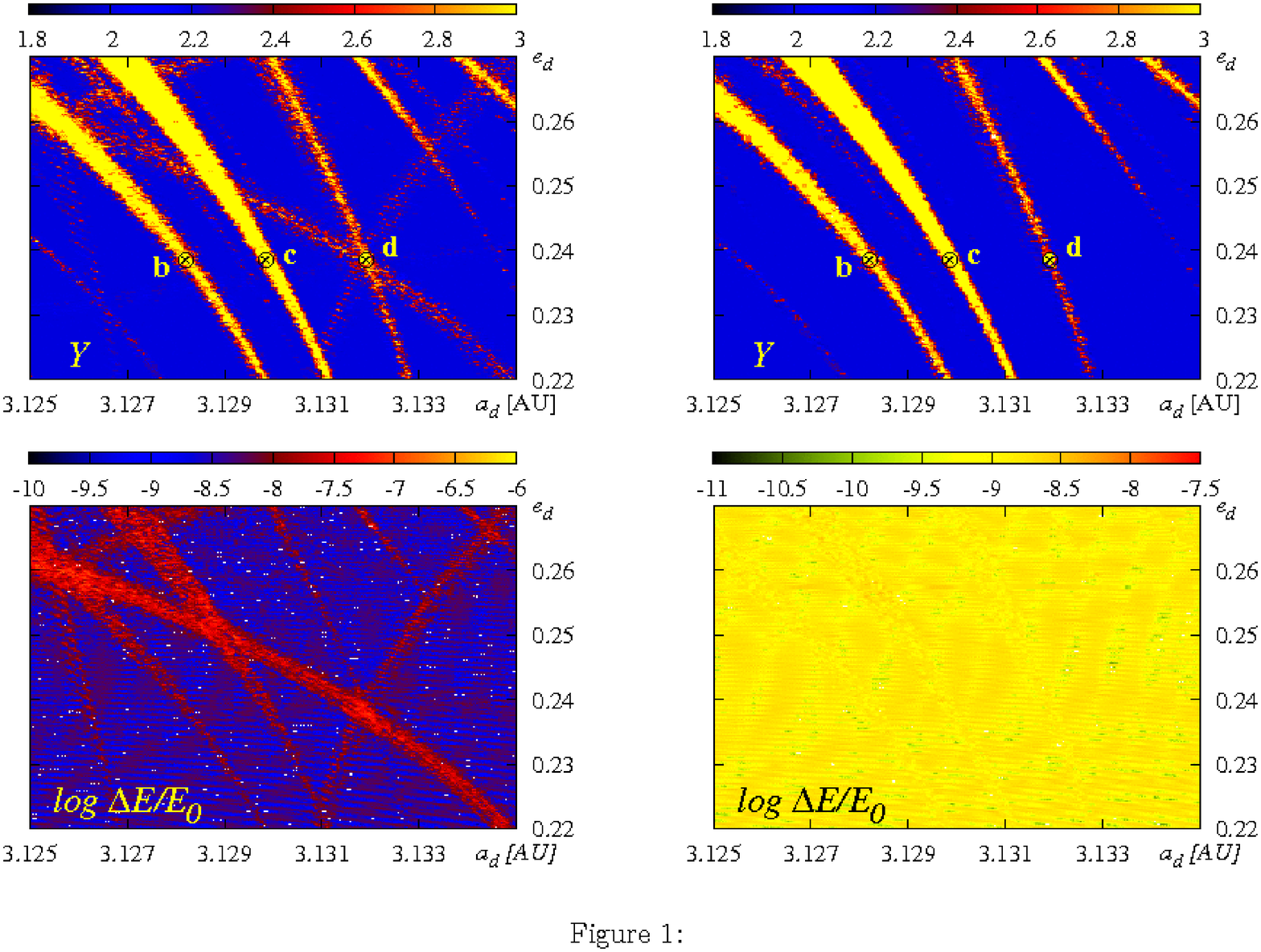}
}
\caption{
Close-up's of the MEGNO dynamical map shown in Fig.~\ref{fig:fig4} within
rectangle labeled with III, illustrating the fine structure of the phase space.
The integration time  $\sim 3\times 10^5$~yr is equivalent to $\sim 4.8 \cdot
10^4$ orbital periods of the putative outermost planet~d. The resolution is
$200\times 200$ points. Panels in the top are for the MEGNO computed by the
symplectic algorithm: the left panel is for the time-step of 16~days, the right
panel if for the time step of 10~days. The bottom row is for the relatitve error
of the total energy, for the same time steps, respectively.
}
\label{fig:fig6}
\end{figure*}

A closer inspection of the area  III in Fig.~\ref{fig:fig4} reveals a multitude
of weakly unstable solutions. To show such structures in more detail, we
computed a close-up of that area (Fig.~\ref{fig:fig6}). The resolution of the 
maps is  $200\times200$ data points, the integration interval is $\sim3\cdot
10^5~\mathrm{yr}  \approx 5\cdot10^4 P_{\idm{d}}$. The time step at the left
column is 16~days and it provides the relative error of the total energy at the
level of $10^{-8}$ . Apparently, in spite of relatively short integration time,
the map uncovers sophisticated structure of the two-body and three-body
resonances. To identify some of them, we choose initial condition marked by
small crossed circles and labeled in the map with  \textbf{b},   \textbf{c}, and
\textbf{d}, respectively.

For initial condition \textbf{b} we found that
$
n_\mathrm{b} - 11 n_\mathrm{c} + 15 n_\mathrm{d} \approx -0.1^{\circ}/\mbox{yr},
$
i.e., indicating three-body MMR $+1\mbox{b}:-11\mbox{c}:+15\mbox{d}$;
for initial condition  \textbf{c}  we have got
$
10 n_\mathrm{c} - 27 n_\mathrm{d} \approx -0.3^{\circ}/\mbox{yr},
$
corresponding to the $+10\mbox{c}:-27\mbox{d}$ MMR of the outer
giants; and for initial condition \textbf{d}  we have
$
n_\mathrm{b} - n_\mathrm{c} -12 n_\mathrm{d} \approx -0.2^{\circ}/\mbox{yr},
$
indicating the three-body $+1\mbox{b}:-1\mbox{c}:-12\mbox{d}$ MMR, respectively.
All these resonances excite chaotic configurations. That can be demonstrated
through observing the time evolution of their critical angles (see
Fig.\ref{fig:fig8}b,c,d). In all those instances, we found that the libration
alternates with circulation of these angles, hence confirming that the relevant
configurations are close to the resonance separatrices.

A particularly interesting star-like structure can be be seen around the
initial condition  \textbf{d} (the upper-left panel in Fig.~\ref{fig:fig6}).
In that area, the two-body $10\mbox{c}:27\mbox{d}$
MMR and many week three-body resonances  are active, for instance,
$ +1\mbox{b}:-11\mbox{c}:+15\mbox{d} \approx -1.35^{\circ}/\mbox{yr}$,
$+1\mbox{b}:-1\mbox{c}:-12\mbox{d} \approx -0.2^{\circ}/\mbox{yr}$,
$+2\mbox{b}:-12\mbox{c}:+3\mbox{d} \approx -1.5^{\circ}/\mbox{yr}$,
$+3\mbox{b}:-13\mbox{c}:-9\mbox{d} \approx -1.7^{\circ}/\mbox{yr}$, and
$+10\mbox{c}:-27\mbox{d} \approx 1.2^{\circ}/\mbox{yr}$.

\begin{figure}
\centerline{
  \includegraphics[width=7cm]{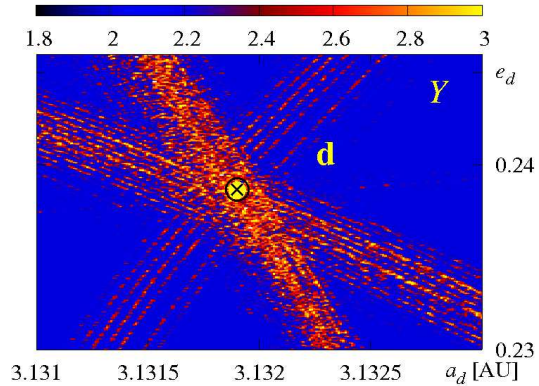}
}
\caption{
Close-up's of the MEGNO dynamical map shown in the left panel of
Fig.~\ref{fig:fig6} The integration time  $\sim 2\times 10^5$~yr is equivalent
to $\sim 3.2 \cdot 10^4$ orbital periods of the putative outermost planet~d. The
resolution is $200\times 480$ points. The MEGNO is computed by the symplectic
algorithm with the time-step of 16~days.
}
\label{fig:fig7}
\end{figure}

\begin{figure*}
\centerline{
  \includegraphics[width=5.2in]{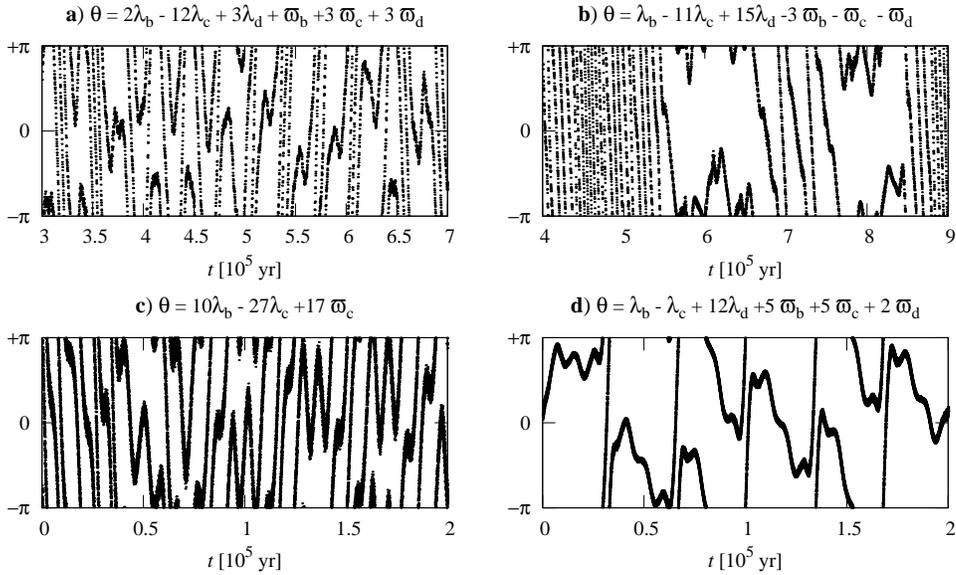}
}
\caption{
Time-evolution of the critical arguments of some resonances illustrated in the
dynamical maps in Figs.~\ref{fig:fig4} and~\ref{fig:fig6}. The panels are
labeled accordingly with initial conditions marked by small crossed circles in
these  dynamical maps. See the text for more details.
}
\label{fig:fig8}
\end{figure*}

One might be tempted to attribute this sophisticated structure to the so called
Arnold web \citep{Cincotta2002}.  Indeed, a closer look at the branches of the
web shows new fine details and extremely complex dynamical structure, in that
zone, illustrated in the close-up map around initial condition  \textbf{d},
Fig.~\ref{fig:fig7}.   But the truth is that this particular structure is mostly
spurious, and it occurred due to an improper choice of the integration step.  To
shed more light on that issue, we show the map of the relative error of the
total energy (the left-bottom panel in Fig.~\ref{fig:fig6}).  The coincidence of
higher energy error streaks (bottom left) with instability patterns detected by
MEGNO (top left) is not conclusive by itself, but when we recompute both maps
using a smaller time step of 10~days (panels in the right column of
Fig.~\ref{fig:fig6}), we notice that the web patterns of higher MEGNO disappear
(Fig.~\ref{fig:fig6}, top right) and the energy error map significantly flattens
(Fig.~\ref{fig:fig6}, bottom right). We conclude that two additional resonance
lines that crossed at \textbf{d} were generated by the so-called `step-size
resonances' \citep{Rauch1999} between proper frequencies of the system and the
sampling frequency of the constant step integrator. The effect of step-size
resonance in a constant step integrator can be avoided either by using a
sufficiently small integration step or by the application of high-order schemes.
Unfortunately, both approaches lead to more time consuming algorithms.

As we could expect,  the symplectic scheme outperforms the classical integration
algorithms. For instance, we found the the MEGNO code driven by the
Bulirsh-Gragg-Stoer ODEX  integrator \citep{Hairer1995} with the relative
accuracy set to $10^{-13}$ requires a similar CPU time, as the Laskar-Robutel
SBAB3 scheme with 4~days step-size, but the former leads to a much larger,
secularly growing energy error (which is larger by 2-3 orders of magnitude).

\section{Conclusions}

The use of Poincar\'e variables in the studies of  the dynamics of close-to
integrable planetary systems offers many advantages. The variables are canonical
and offer a simple form of a reduced Hamiltonian. The Hamiltonian can be split
into a sum of two separately integrable parts: the Keplerian term and a small
perturbation. As such, it can serve to construct a symplectic integrator based
on any modern composition method, including the recent ones invented by
\citet{LasRob}. The tangent map computed with the same integration scheme
provides an efficient way of computing the estimate of maximal Lyapunov exponent
in terms of relatively recent fast indicator MEGNO. The method proves to be much
more efficient than general purpose integrators (like the Bulirsh-Stoer-Gragg
method). Besides, it provides the conservation of the integrals of energy and
the angular momentum that is crucial for resolving the fine structure of the
phase space. From the practical point of view, the symplectic algorithms are
relatively simple for numerical implementation.

Using the numerical tools, we investigate the long term stability of extrasolar
planetary system hosted by HD~37124. The orbital parameters in the set of our
best, self-consistent Newtonian fits \citep{Gozdziewski2006a}  are in accord
with the discovery paper \citep{Vogt2005}. Nevertheless, the observational
window of the system is still narrow and the derivation of the model consistent
with observations is difficult and, in fact, uncertain. The dynamical maps
reveal that the relevant region of the phase space, in the neighborhood  of the
mathematically best fit, is a strongly chaotic and unstable zone. The fitting
algorithm (GAMP) that relies on eliminating strongly unstable fits founds
solutions with a similar quality [in terms of $\Chi$] that yields the formal
solution. Moreover, they are shifted towards  larger semi-major axes and much
smaller eccentricities  of the outermost planet. The orbital evolution of two
outer planets is confined to a zone spanned by a number of low-order two-body
and three-body MMRs. In particular, the three-body MMRs may induce very unstable
behaviors that manifest themselves after many Myrs of an apparently stable and
bounded evolution. To deal with such a problem, the stability of the best fits
should be examined over a time-scale that is much longer than the one required
when only the two-body MMRs are considered. In accord with the  dynamical maps,
the  stable fits to the RV of HD~37124 should have small eccentricity of the
outermost  planet~d, not larger than 0.2-0.3. Moreover, the stable
configurations of the HD~37124 system are puzzling.  The best-fit mathematical
three-planet model is surprisingly distant, in the phase space of initial
conditions, from the zone of stable solutions consistent with the RV. It remains
possible that other bodies are present in the system and the three-planet model
is  not adequate to explain the RV variability, in spite that it provides 
apparently perfect fits. Yet, isolated initial conditions or even sets of
best-fit solutions do not provide a complete answer on the system configuration.
Then the fast indicator approach is essential and helpful to resolve the
dynamical structure of the phase space.

The results of our experiments confirm and warn that all numerical methods should be
applied with great care.  All symplectic methods are constant step integrators.
In that case one should be cautious about the possibility of generating spurious
resonance webs. A proper way to avoid them is to repeat computations with a
different integration step in order to detect step-dependent patterns.

\section{Acknowledgments}
We thank David Nesvorn{\'y} for a review and comments that improved that
manuscript. This work is supported by KBN Grant No. 1P03D-021-29.

\bibliographystyle{mn2e}
\bibliography{biblio}
\end{document}